\documentclass[prb,twocolumn,superscriptaddress,showpacs,%
preprintnumbers,amsmath,amssymb]{revtex4}
\usepackage{amsmath}
\usepackage{graphicx}
\usepackage{dcolumn}

\begin{document}

\title{Braess paradox at the mesoscopic scale}

\author{A. A. Sousa}\email{ariel@fisica.ufc.br}
\affiliation{Departamento de F\'isica, Universidade Federal do
Cear\'a, Caixa Postal 6030, Campus do Pici, 60455-900 Fortaleza,
Cear\'a, Brazil}
\affiliation{Department of Physics, University of
Antwerp, Groenenborgerlaan 171, B-2020 Antwerp,
Belgium}
\author{Andrey Chaves}\email{andrey@fisica.ufc.br}
\affiliation{Departamento de F\'isica, Universidade Federal do
Cear\'a, Caixa Postal 6030, Campus do Pici, 60455-900 Fortaleza,
Cear\'a, Brazil}
\author{G. A. Farias}
\affiliation{Departamento de F\'isica, Universidade Federal do
Cear\'a, Caixa Postal 6030, Campus do Pici, 60455-900 Fortaleza,
Cear\'a, Brazil}
\author{F. M. Peeters}\email{francois.peeters@ua.ac.be}
\affiliation{Department of Physics, University of
Antwerp, Groenenborgerlaan 171, B-2020 Antwerp,
Belgium}
\affiliation{Departamento de F\'isica, Universidade
Federal do Cear\'a, Caixa Postal 6030, Campus do Pici, 60455-900
Fortaleza, Cear\'a, Brazil} 

\date{ \today }

\begin{abstract}
We theoretically demonstrate that the transport inefficiency recently found experimentally for branched-out mesoscopic networks can also be observed in a quantum ring of finite width with an attached central horizontal branch. This is done by investigating the time evolution of an electron wave packet in such a system. Our numerical results show that the conductivity of the ring does not necessary improves if one adds an extra channel. This ensures that there exists a quantum analogue of the Braess Paradox, originating from quantum scattering and interference.
\end{abstract}


\maketitle

\section{Introduction}

Suppose that two points A and B of a network are connected only by two possible paths (e.g. roads in a traffic network, or wires in an electricity network). One would intuitively expect that adding to the network a third path connecting these two points would lead to an improvement of the flux through the pre-existing roads and, consequently, to a transmission enhancement. However, the so-called Braess paradox \cite{Braess, Cohen1, Lin} of games theory states that this is not necessarily the case: under specific conditions, \cite{Xia} adding a third path to a network may lead to transport inefficiency instead. This effect has been even observed in traffic networks in big cities, where \textit{closing} roads \textit{improves} the flux in traffic jams, \cite{1} or in electricity networks, where it has been demonstrated that adding extra power lines may lead to power outage, due to desynchronization. \cite{Cohen,2,3}

A recent paper \cite{Huant} showed both experimental and theoretical evidence of a very similar effect, but on a mesoscopic scale: they observed that branching out a mesoscopic network does not always improve the electrons conductance through the system. As they were dealing with a system consisting of wide transmission channels, quantum interference effects are not expected to be relevant. \cite{Huant2}

In this paper, we demonstrate that the transport inefficiency in branched out devices also occurs on a nano scale, when only few sub-bands are involved, and transport is strongly influenced by quantum effects. For this purpose, we investigate wave packet propagation through a circular quantum ring attached to input (left) and output (right) leads, \cite{Chaves} in the presence of an extra channel passing diametrically through the ring. Our results demonstrate that increasing the extra channel width does not necessarily improve the overall current. The fundamental reasons behind this effect, which are related to quantum scattering and interference, are discussed in details in the following Sections.

\section{Theoretical model}

\begin{figure}
\begin{center}
\includegraphics[width=1.0\linewidth]{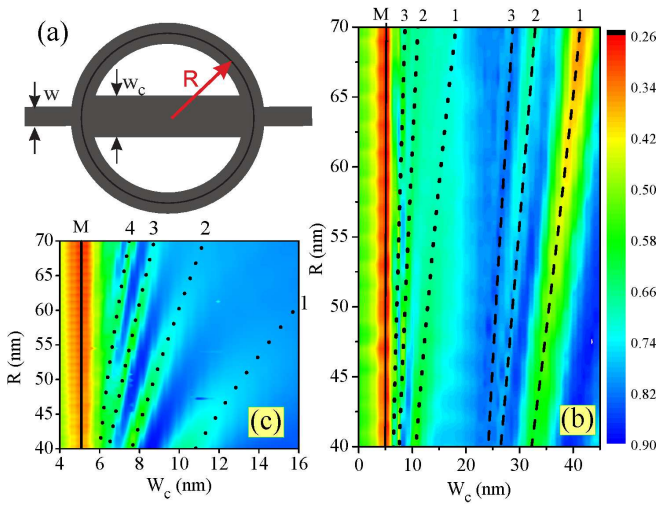}
\end{center}
\caption{ \label{fig1} (a) Sketch of the system under investigation: a quantum ring with average radius $R$, attached to input (left) and output (right) channels with the same width as the ring ($W = 10$ $nm$), and to an extra horizontal channel of width $W_c$. (b) Contour plots of the transmission probabilities as a function of the extra channel width and ring radius. The solid, dashed and dotted lines indicate seven minima that are discussed in the text. A zoom of the $4$ nm $< W_c < 16$ nm region with the logarithm of the transmission is shown in (c).}
\end{figure}

We consider an electron confined in a circular quantum ring attached to input (left) and output (right) leads, \cite{Chaves} in the presence of an extra channel passing diametrically through the ring, as sketched in Fig. \ref{fig1}(a). Both the ring and the leads are assumed to have the same width $W = 10$ nm, whereas different values of the extra channel width $W_c$ are considered.

As initial wave packet, we consider a plane wave with wave vector $k_0 = \sqrt{2m_e\epsilon}/\hbar$, where $\epsilon$ is the energy and $m_e$ is the electron effective mass, multiplied by a Gaussian function in the $x$-direction, and by the ground state $\phi_0 (y)$ of the input channel in the $y$-direction,
\begin{equation}\label{eq.init}
\Psi (x,y,0) = \exp\left[ik_0x - \frac{(x-x_0)^2}{2\sigma_x^2}\right]\phi_0(y).
\end{equation}
Several papers have reported calculations on wave packet propagation in nanostructured systems, \cite{Saskia, Kramer, Fehske, Szafran} hence, a number of numerical techniques for this kind of calculation is available in the literature, such as the expansion of the time evolution operator in Chebyshev polynomials, \cite{Fehske2} and Crank-Nicolson based techniques. \cite{Szafran2} In the present work, the propagation of the wave packet in Eq. (\ref{eq.init}) is calculated by using the split-operator technique \cite{Chaves, Splitoperator, Splitoperator2} to perform successive applications of the time-evolution operator, i.e. $\Psi(x,y,t+\Delta t) = \exp\left[-iH\Delta t/\hbar\right]\Psi(x,y,t)$, where $\Delta t$ is the time step. The Hamiltonian $H$ is written within the effective mass approximation, describing an electron constrained to move in the $(x,y)$-plane and confined, by external potential barriers of height $V_0$, to move inside the nanostructured region represented in gray in Fig. \ref{fig1}(a), where the potential is set to zero. The interface between the confinement region and the potential barrier is assumed to be abrupt. Nevertheless, considering smooth potential barriers would not affect the qualitative behavior of the results to be presented here, since the effect of such smooth interfaces has been demonstrated to be mainly a shift on the eigenenergies of the system. \cite{Wang, King} The $(x,y)$-plane is discretized in a $\Delta x = \Delta y = 0.4$ nm grid, and the finite differences technique is used to perform the derivatives coming from the kinetic energy terms of the Hamiltonian. Imaginary potentials \cite{ImaginaryBoundary} are placed on the edges of the input and output channels, in order to absorb the propagated wave packet and avoid spurious reflection at the boundaries of the computational box. As the wave packet propagates, we compute the probability density currents at the input and output leads, which, when integrated in time, gives us the reflection and transmission probabilities, respectively, from which the conductance can be calculated.

As the fabrication of InGaAs quantum ring structures have already been reported in the literature, \cite{Hackens} we assume that the ring, channel and leads in our model are made out of this material, so that the electron effective mass is taken as $m_e = 0.041 m_0$. Nevertheless, the qualitative features of the results presented in the following Section does not depend on specific material parameters.


\section{Results and discussion}

Contour plots of the calculated transmission probabilities are shown in Fig. \ref{fig1}(b) as a function of the ring radius $R$ and the width $W_c$ of the extra channel. Notice that the extra channel in the system is opened in the horizontal direction, namely, parallel to the input and output leads, being practically just a continuation of these leads. Even so, instead of improving the transmission, the existence of such a channel surprisingly \textit{reduces} the transmission probability for specific values of $W_c$, leading to several minima in each curve. In what follows, we discuss the origin of several of these minima, indicated by the solid, dashed and dotted curves in Fig. 1(b). 

The position of the minima labeled as 1, 2 and 3 in Figs. \ref{fig1} (b,c) changes with the ring radius, which indicates that these minima are related to a path difference, i.e. to an interference effect. Let us provide other arguments to support this indication: in a very simplistic model, consider that part of the wave packet travels through the central channel, while the other part passes through the ring arms. The latter runs a length $\approx \pi R$ while going from the input to output leads, whereas the former runs through the $2R$ diameter of the ring. The condition for destructive interference is:
\begin{equation}\label{eq.inter}
\gamma\frac{\pi R}{\lambda} - \frac{2 R}{\bar{\lambda}} = n+\frac{1}{2},
\end{equation}
where $\lambda = 2 \pi\big/ \sqrt{2m_e\epsilon/\hbar^2}$ ($\bar{\lambda} = 2 \pi\big/\sqrt{2m_e(E-\bar{E_j})/\hbar^2}$) is the wave length in the ring arms (extra channel), $E_i$ ($\bar{E_j}$) is the energy of the $i$-th ($j$-th) eigenstate of the input lead (extra channel), and $E = \epsilon + E_i$ is the total energy of the wave packet. The parameter $\gamma$ is close to one and accounts for the fact that the effective arm length may be slightly different from $\pi R$ [see Fig. 1(a)]. By substituting these expressions for $\lambda$ and $\bar{\lambda}$ in Eq. (\ref{eq.inter}), one obtains
\begin{equation}\label{eq.conditioninter}
\bar{E_j} = E - \frac{\hbar^2\pi^2}{2m_e}\left[\frac{\gamma}{2}\sqrt{\frac{2m_e}{\hbar^2}\epsilon}-\left(n+\frac{1}{2}\right)\frac{1}{R}\right]^2,
\end{equation}
Hence, this equation gives the condition for the interference related minima in the transmission probability. The extra channel eigenstates $\bar{E_j}$ depend on $W_c$ - which can be fairly well approximated by $\bar{E_j} \simeq \beta/W_c^{1.85}$ for large $W_c$ (notice that the structure has finite potential barriers, therefore, the infinite square well relation $\bar{E_j} \propto 1/W_c^2$, is no longer valid). Therefore, the minima for large $W_c$ are expected to occur for 
\begin{equation}\label{eq.conditioninterW}
W_c^{(n)} = \left\lbrace\frac{\beta}{E - \frac{\hbar^2\pi^2}{2m_e}\left[\frac{\gamma}{2}\sqrt{\frac{2m_e}{\hbar^2}\epsilon}-\left(n+\frac{1}{2}\right)\frac{1}{R}\right]^2}\right\rbrace^{1/1.85},
\end{equation}
which are shown in Fig. \ref{fig1}(b) by black dashed lines for $n = 1, 2$ and $3$. The model fits very well the numerically obtained positions for these minima for $\gamma = 0.865$. The $n = 0$ minimum occurs outside of the investigated range $W_c$. The wave packet in this case has total energy $E = 124$ meV, with $\epsilon = 70$ meV and $E_0 = 54$ meV (ground state of the $W = 10$ nm input lead). For the 26 nm $< W_c < $ 42 nm range in Fig. \ref{fig1}(b), the eigenstates of the channel, which are accessible by the electron with this energy, are the ground state and the second excited state. The first and third excited states, although still having energy lower than 124 meV for this range of $W_c$, are not accessible by the wave packet because of the even symmetry of the initial wave packet with respect to the $x$-axis, while these excited states of the channel are odd. Therefore, the part of the wave packet that goes through the central channel under these conditions populates mostly the second excited state, but has also some projection on the ground state and none on the other states. The fitting of $\bar{E_j}$ for the second excited state ($j$ = 2) has $\beta \approx 3000$ meV nm$^{1.85}$, which is the value used in Eq. (\ref{eq.conditioninterW}) to obtain the dashed curves in Fig. \ref{fig1}(b). 

The $n = 1, 2$ and $3$ minima occurring for 7 nm $< W_c <$ 15 nm in Fig. \ref{fig1}(b) can also be obtained from Eq. (\ref{eq.conditioninterW}) but, since this is a lower $W_c$ range, the dependence of $\bar{E_j}$ on $W_c$ will have a different exponent, one needs to replace 1.85 by 1.50 in Eq. (\ref{eq.conditioninterW}). Besides, for such low $W_c$, the wave function travels predominantly through the ground state sub-band of the extra channel, so that one must consider the $j = 0$ state of this channel, which has $\beta \approx 56.99$ meV nm$^{1.50}$ in this range. The results for this model are shown as black dotted lines in Fig. \ref{fig1}(b). To show these minima more clearly, we present in Fig. \ref{fig1} (c), a magnification of the logarithm of the probability in the low $W_c$ region. The numerically obtained minima are well fitted by the model of Eq. (\ref{eq.conditioninterW}) for $\gamma = 0.925$ with $n = 1, 2, ... 4$ (see dotted lines).


\begin{figure}
\begin{center}
\includegraphics[width=0.8\linewidth]{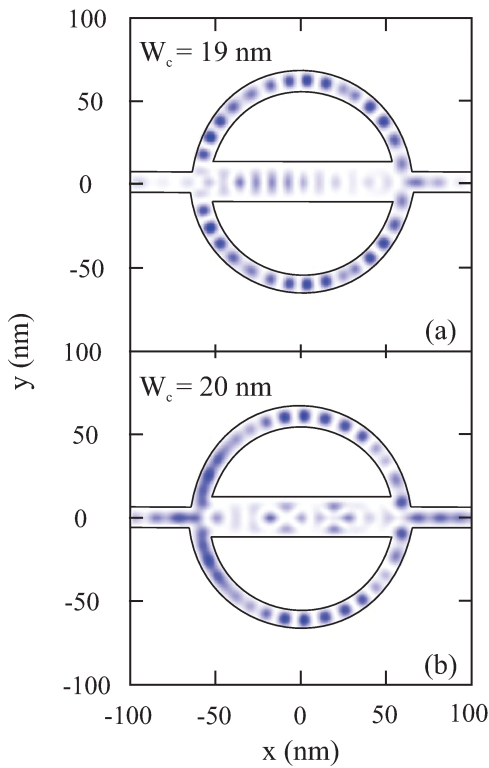}
\end{center}
\caption{\label{waveproj} Snapshot of the propagating wave function at $t = 900$ fs for two values of the extra channel width: 19 nm (a) and 20 nm (b).}
\end{figure}


In order to demonstrate that for lower (higher) values of $W_c$ the wave function inside the extra channel is predominantly in its ground (second excited) state, Fig. \ref{waveproj} shows a snapshot of the propagating wave function at $t = 900$ fs for two values of the extra channel width: $W_c$ = 19 nm (a) and 20 nm (b). In the former case, the wave function inside the extra channel exhibits predominantly a single maximum peak around $y$ = 0, which suggests a large contribution of the ground state eigenfunction in the wave packet within this region. Similar results are obtained for lower values of the channel width $W_c$. However, the results for a slightly larger $W_c$ = 20 nm are qualitatively different, exhibiting three peaks along the $y$-direction inside the extra channel, which implies a higher contribution of the second excited state on the wave function in this region. 


\begin{figure}
\begin{center}
\includegraphics[width=1\linewidth]{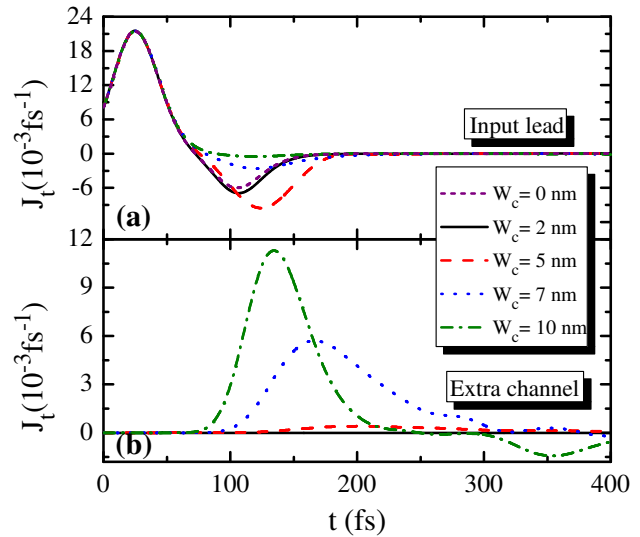}
\end{center}
\caption{ \label{fig2} Probability density currents as a function of time, calculated (a) in the input lead and (b) in the extra channel, for different values of the extra channel width $W_c$, for wave packet energy $\epsilon = 70$ meV.}
\end{figure}


Differently from the other minima, the position of the first minimum $M$ in Fig. \ref{fig1}(b) appears around $W_c = 5$ nm and does not change with the radius $R$. Therefore, this minimum cannot be related to the above discussed interference effect. In order to understand the origin of the $M$ minima, we show in Fig. \ref{fig2} the integrated current $J_t$ in the input lead and the extra channel. Fig. \ref{fig2}(a) exhibits a high negative peak for $W_c = 2$ nm and 5 nm at $\approx 100$ fs and $\approx 140$ fs, respectively, which represents a strong reflection of the wave packet at the ring - channel junction. \cite{footnote} This is confirmed by the very low currents observed for these cases inside the extra channel, in Fig. \ref{fig2}(b). On the other hand, for $W_c = 7$ nm the reflection peak in the input lead becomes very weak, while for $W_c =$ 10 nm, almost no reflection is observed. For the latter two cases instead, large current peaks are observed inside the extra channel. This is a clear indication that the transmission inefficiency in the low $W_c$ case is not related to interference effects, but rather to scattering at the ring-channel junction, since the wave packet barely enters the extra channel when it is too narrow.

\begin{figure}
\begin{center}
\includegraphics[width=1.0\linewidth]{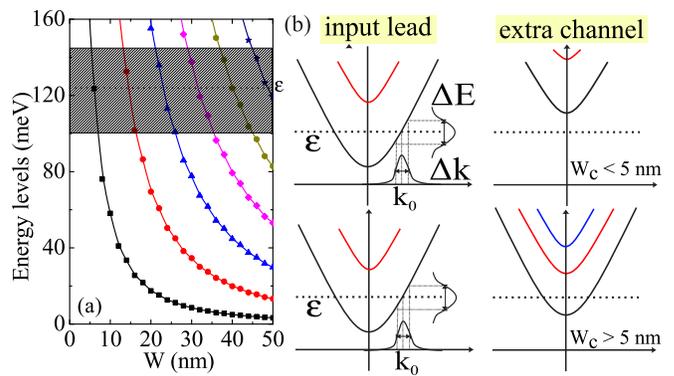}
\end{center}
\caption{\label{fig3} (a) Eigenstates of a finite quantum well as a function of its width. (b) Diagram representing the energy sub-bands in the input lead and in the extra channel. The horizontal dotted line is the average energy of the wave packet used in Figs. \ref{fig1} and \ref{fig2}, and the shaded area in (a) illustrates the FWHM of the energy distribution of this wave packet.}
\end{figure}

We discuss now the possibility of having part of the incoming wave packet passing through a narrow extra channel. Both the leads and the extra channel have discrete eigenstates due to the quantum well confinement in the $y$-direction, whose energy levels are shown in Fig. \ref{fig3}(a) as a function of the well width. In the $x$-direction, parabolic sub-bands stem from these eigenstates, as illustrated in Fig. \ref{fig3}(b). The incoming wave packet considered in Figs. \ref{fig1} - \ref{fig2} has $\epsilon = 70$ meV on top of its ground state energy in the input lead, $E_0 = 54$ meV (for $W = 10$ nm). This energy is represented by the dotted horizontal lines in Figs. \ref{fig3}(a) and (b). The wave packet has a Gaussian distribution of energies of width $\Delta E = \hbar^2/m_e k_0\Delta k$, where $\Delta k = 2\sqrt{ln2}/\sigma_x$ is the full width at half maximum (FWHM) of the wave vector distribution, which is represented by the shaded area around the dotted line in Fig. \ref{fig3}(a). A narrow extra channel has a very high ground state sub-band energy, so that no component of the incoming wave packet energy has enough energy to pass through the channel. As the extra channel width $W_c$ increases, its sub-band energies decrease, allowing the incoming wave packet to travel through this channel. These two situations are illustrated in the upper and lower figures of Fig. \ref{fig3}(b), respectively. Notice that the upper boundary of the energy distribution (shaded area) in Fig. \ref{fig3}(a) is crossed by the second excited state energy curve (blue triangles) approximately at $W = 20$ nm. This explains the drastic difference between the wave functions within the extra channel with $W_c$ = 19 nm and 20 nm, observed in Fig. \ref{waveproj}: in the latter case, the wave function has a significantly larger part of its energy distribution above the second excited state energy, allowing it to have a larger projection on this state. 

Therefore, the counter-intuitive result observed in Figs. \ref{fig1}, namely, the transmission reduction as the extra channel width increases for lower values of $W_c$, is a pure quantum scattering effect. For classical particles, such an extra channel with any width would allow the passage of the particles and, consequently, improve the transmission. However, a quantum channel has a confinement energy (ground state) and, if the energy of the incoming particle is lower than this minimum, the particle is not allowed to pass through the channel. Therefore, adding a narrow extra channel to the system, which effectively also adds extra scattering, does not add an extra path for the wave packet, because of the very high ground state energy of the narrow channel. This mechanism, which is illustrated by the band diagrams in Fig. \ref{fig3}(b), leads to the strong reflections observed in Fig. \ref{fig2} for $W_c = 2$ nm and 5 nm. For $W_c > 5$ nm, a significant part of the $E = 70$ meV wave packet has enough energy to go through the extra channel, explaining the increasing transmission as $W_c$ increases above 5 nm. This also suggests that incoming wave packets with higher energy would need lower extra channel widths to pass, which is indeed observed, as we will discuss further on.


\begin{figure}
\begin{center}
\includegraphics[width=1.0\linewidth]{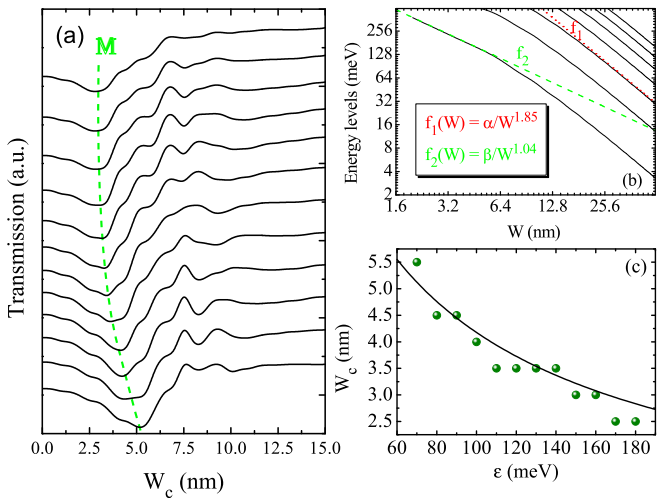}
\end{center}
\caption{\label{fig5} (a) Transmission probabilities as a function of the extra channel width in the vicinity of the minimum labeled as $M$ in Fig. \ref{fig1}(a), for several values of the wave packet energy $\epsilon$ = 70 (bottom curve), 80, ... 180 meV (top curve). The curves were shifted 0.1 up from each other, in order to help visualization. (b) Energy levels (solid) as a function of the channel width, plotted in a log scale, along with two fitting functions (dashed curves), for large ($f_1$) and small ($f_2$) values of the channel width. (c) Numerically obtained (symbols) positions of the $M$ minima as a function of the wave packet energy, along with the results (solid curve) of the analytical model, given by Eq. (\ref{eq.minimaM1}).}
\end{figure}

In fact, the position of $M$ strongly depends on the wave packet energy, as shown in Fig. \ref{fig5}(a), where the transmission probability in the vicinity of $M$ is plotted as a function of the channel width $W_c$ for several values of the energy, ranging from 70 meV (bottom curve) to 120 meV (top curve), with 10 meV intervals. The ring radius is fixed as $R = 60$ nm, and each consecutive curve in this figure is shifted by 0.1. If the energy dependence of the position of $M$ is due to the above discussed quantum effect, it should be possible to predict the position of these minima from the following argument: the highest energy components of the wave packet have energy around $\approx E+\Delta E/2$. These components would be allowed to pass through the extra channel, consequently improving the current, provided the channel width is wide enough to have a ground state energy as low as their energy, i.e. if $\bar{E_0} < E + \Delta E/2$. For low values of $W_c$, the ground state energy of the channel is well approximated by $\bar{E_0} = \alpha/W_c^{1.04}$, for $\alpha = 8.65$ eV, as shown by the green dashed line ($f_2(W)$ function) in Fig. \ref{fig5}(b). Notice it is a different power from the one used in Eq. (\ref{eq.conditioninterW}), which is valid only for higher $W_c$ values. The red dotted line ($f_1(W)$ function) in Fig. \ref{fig5}(b) is an example of fitting for high values of $W_c$, which was used in Eq. (\ref{eq.conditioninterW}). Figure \ref{fig5}(b) is in log-log scale, so that the power laws in $f_1(W)$ and $f_2(W)$ are shown as straight curves, whose slopes are the functions' exponents. Using this expression for $\bar{E_j}$, one obtains the following approximate expression for the position of the $M$ minima
\begin{equation}\label{eq.minimaM1}
W_c^{(M)} = \frac{6103}{\left(\epsilon + E_0 + \hbar\sqrt{\frac{\epsilon}{2m_e}}\Delta k\right)^{1/1.04}},
\end{equation}
which is shown by the solid curve in Fig. \ref{fig5}(c). Notice the rather good agreement with the numerically obtained positions of the $M$ minima, represented by the symbols. 

It is important to point out that the exponents 1.85, 1.50 and 1.04, as well as the values of $\alpha$ and $\beta$, found for the fitting functions for the eigenstate energies as a function of the well width and used in Eqs. (\ref{eq.conditioninterW}) and (\ref{eq.minimaM1}), were obtained for an abrupt interface between the potential barriers and the confining region. These values must be slightly modified in the case of smooth potential barriers.


\begin{figure}
\begin{center}
\includegraphics[width=0.8\linewidth]{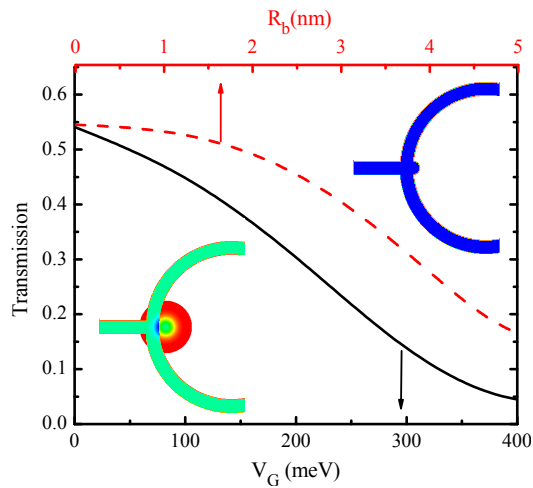}
\end{center}
\caption{\label{fig6} Transmission probabilities for a $\epsilon$ = 70 meV wave packet scattered by two kinds of defects in the lead-ring junction: a Gaussian attractive potential of depth $V_G$ (solid, bottom axis) and width $\sigma_G = 5$ nm, and a circular bump of radius $R_b$ (dashed, top axis), which are schematically illustrated in the lower and upper insets, respectively.}
\end{figure}

Our results, therefore, demonstrate that the $M$ minima in Figs. \ref{fig1} and \ref{fig5} are a consequence of a competition between two effects: (i) the quantum scattering in the ring-channel junction, which increases the reflection when a narrow extra channel is added, and (ii) the improvement in the transmission resulting from the part of the wave packet that has enough energy to propagate through the sub-bands of the extra channel. The former suggests that adding extra scatterers at the input lead-ring junction leads to a larger reflection back into the input lead. In order to verify this, we consider two situations that mimic the appearance of an extra ``blind" channel (see insets of Fig. \ref{fig6}): one is the presence of an attractive Gaussian potential \cite{Chaves} $V_a(x,y) = -V_G\exp\left\{[(x-x_g)^2 + y^2]/2\sigma_G^2\right\}$ close to the lead-ring junction, and the other is a circular bump of radius $R_b$ in the inner boundary of the ring. Fig. \ref{fig6} shows the transmission probabilities for $\epsilon = 70$ meV as a function of the Gaussian potential depth $V_G$ (bottom axis) and the radius $R_b$ (top axis) of the circular bump. In both cases, the transmission is reduced in the presence of the extra scatterer, which supports the idea that the transmission reduction in the low $W_c$ range in Figs. \ref{fig1} and \ref{fig5} is indeed a consequence of extra scattering created by the opening of the extra channel, which is however effectively blind, since the bottom of the ground state sub-band of the a narrow channel has energy higher than that of the incoming electron wave packet. 

All the results in this work were calculated for sharp connections between the ring, the extra channel, and the input and output leads. However, qualitatively similar results are also obtained for smooth junctions \cite{Chaves} between these parts of the system. Moreover, different ring geometries would shift the high $W_c$ minima, which are related to quantum interference, by effectively changing the electronic paths, while impurities in the ring could suppress these minima, by destroying phase coherence. However, neither impurities nor different ring geometries can affect the low $W_c$ minimum ($M$), since it is related only to quantum scattering in the input lead-ring junction, which does not depend on these features.

The original version of the Braess paradox, described in details in Ref. \onlinecite{Braess}, discusses how the travel time between two points connected by only two possible roads, A and B, changes if these two roads are inter-connected by a third road C. If one considers that the traffic at specific parts of A and B depend on the number of drivers in these roads, then, depending on the (partial) travel time through this new connection C, the dominant strategy turns out to consist in starting in one road and changing to the other road through the connection C, and therefore, all players (drivers) would take this path. This strategy, though leading to the Nash equilibrium situation of this system, represents an increase in the travel time - lower travel times could even be reached if the drivers agree not to use the connection C \textit{a priori}, but in a scenario of selfish drivers, they would switch roads until the equilibrium is reached, despite the reduction in overall performance. Therefore, the classical Braess paradox is closely related to an unsuccessful attempt to optimize the travel time through a traffic network by the drivers. The transport properties of the branched out mesoscopic network investigated in Refs. \cite{Huant, Huant2} is reminiscent of those of the roads network in the original Braess paper just in the sense that it exhibits a reduced overall current when an extra channel is added to the network, depending on the channel width. However, the fundamental reason behind this phenomenon is not clear in Refs. \cite{Huant, Huant2} - it cannot be an interference effect, since this is not a coherent system, but it is not also due to an optimization of the currents, as in the classical paradox, since the model in these papers does not involve non-linear equations or iterative calculations of the overall current flow. On the other hand, for the quantum case investigated here, where such a transmission reduction in the presence of an extra channel is also observed, the main reason behind this Braess-like paradoxical behavior is quite clear: for small values of the channel width, it is due to quantum scattering effects at the ring-channel junction, whereas for larger widths, it is due to interference effects. Therefore, if one includes the transmission reduction phenomena described here into the category of analogs of Braess paradox, one must keep in mind that, just like most of the other analogs suggested in the literature \cite{Cohen, 2, 3, Huant}, although presenting results similar to those of the original Braess network, in the sense that more paths leads to reduced performance, the reason behind this reduction is not related to an attempt to optimize the flux, but to other fundamental physical properties of the investigated system. 


\section{Conclusions}

We have investigated the propagation of a wave packet through a quantum ring with an extra channel along its diameter. Surprisingly, our results demonstrate that even when an extra channel is added in the horizontal direction, as a continuation of the input and output leads, the transmission through the whole system can be lower than in the absence of such an extra current path. This is evidence of the ''Braess paradox analog'' observed recently for mesoscopic networks. Nevertheless, while the original Braess paradox in games theory is explained in terms of an attempt to optimize the flux, which eventually leads to transport inefficiency in the equilibrium situation, the transport inefficiency observed for the wave packet propagation in quantum systems originates from two possible effects: (i) the quantum scattering of the wave packet in the input channel-ring junction, along with the absence of an allowed energy sub-band for propagation in the central channel when it is too narrow, and (ii) the quantum interference between parts of the wave packets that passed through the central channel and those that propagated through the rings arms.

\acknowledgements

This work was financially supported by PRONEX/CNPq/FUNCAP and the bilateral project CNPq-FWO. Discussions with prof. J. S. Andrade Jr. are gratefully acknowledged. A. A. Sousa has been financially supported by CAPES, under the PDSE contract BEX 7177/13-5.


\begin{references}
\bibitem{footnote} The high positive peak in the input channel is just the incoming wave packet.
\end{references}
\end{document}